\documentclass[useAMS,usenatbib]{mn2e}
\usepackage{amsmath}
\usepackage{graphicx}
\usepackage{multirow}
\usepackage{bm}

\title[Asteroids' Gaia inversion combining lightcurves]{Testing the inversion of asteroids' Gaia photometry combined with ground-based observations}
\author[T. Santana-Ros; P. Bartczak; T. Micha{\l}owski; P. Tanga and A. Cellino]{T. Santana-Ros$^{1}$\thanks{E-mail:
tonsan@amu.edu.pl}; P. Bartczak$^{1}$; T. Micha{\l}owski$^{1}$; P. Tanga$^{2}$ and A. Cellino$^{3}$\\
$^{1} $Astronomical Observatory Institute, Faculty of Physics, Adam Mickiewicz University, S{\l}oneczna 36, 60-286 Pozna{\'n}, Poland\\
$^{2} $Laboratoire Cassiop\'{e}e, Observatoire de la C\^{o}te d'Azur, BP 4229, 06304 Nice Cedex 04, France\\
$^{3} $INAF - Osservatorio Astronomico di Torino, strada Osservatorio 20, I-10025 Pino Torinese (TO), Italy}

\begin{document}

\date{Accepted 2015 March 23. Received 2015 February 9; in original form 2014 December 1}

\pagerange{\pageref{firstpage}--\pageref{lastpage}} \pubyear{2015}

\maketitle

\label{firstpage}

\begin{abstract}
We investigated the reliability of the genetic algorithm which will be used to invert the photometric measurements of asteroids collected by the European Space Agency Gaia mission. To do that, we performed several sets of simulations for 10 000 asteroids having different spin axis orientations, rotational periods and shapes. The observational epochs used for each simulation were extracted from the Gaia mission simulator developed at the Observatoire de la C\^{o}te d'Azur, while the brightness was generated using a Z-buffer standard graphic method. We also explored the influence on the inversion results of contaminating the data set with Gaussian noise with different $\sigma$ values. The research enabled us to determine a correlation between the reliability of the inversion method and the asteroid's pole latitude. In particular, the results are biased for asteroids having quasi-spherical shapes and low pole latitudes. This effect is caused by the low lightcurve amplitude observed under such circumstances, as the periodic signal can be lost in the photometric random noise when both values are comparable, causing the inversion to fail. Such bias might be taken into account when analysing the inversion results, not to mislead it with physical effects such as non-gravitational forces. Finally, we studied what impact on the inversion results has combining a full lightcurve and Gaia photometry collected simultaneously. Using this procedure we have shown that it is possible to reduce the number of wrong solutions for asteroids having less than 50 data points. The latter will be of special importance for planning ground-based observations of asteroids aiming to enhance the scientific impact of Gaia on Solar system science.

\end{abstract}


\section{Introduction}

The potential of the sparse photometric data to provide physical information about asteroids has been extensively proved by several authors (\citealp{cell06}; \citealp{durech07}).  
Generally, the inversion methods used to derive information about the physical properties of asteroids are taking profit of the fact that a simplified version of the asteroids' real shape (triaxial ellipsoid, convex representation) is, in the majority of cases, good enough to describe the asteroid brightness variation due to its rotation for a given period. If the observations are spread over a variety of aspect angles, it is then possible to derive the direction of the asteroid spin axis.

The main challenge to be solved when inverting sparse data is the correct determination of the rotation period. One possible approach to solve this issue is to fit an asteroid spin and shape on a given period interval \citep{kaasa}. Using a convex representation of the asteroid's body shape, some authors have successfully solved the inversion problem for a couple of hundreds of asteroids (\citealp{durech}; \citealp{hanus}). If any \textit{dense} lightcurve is available for the object, the interval is reduced to some range around the observed period, saving a lot of computational time and increasing the solution reliability. But, unfortunately, obtaining full lightcurves of asteroids is a highly time consuming task, thus such observations are actually available only for $\sim$5 000 asteroids (stored in the Minor Planet Lightcurve Database\footnote{http://www.minorplanetcenter.net/}). It is estimated that European Space Agency (ESA) Gaia mission will produce photometric measurements for more than 300 000 asteroids, which means that for the majority of inversion trials the period scanning shall be extended to almost all the possible period values, namely from 2h to several days \citep{eyer}.

Unlike classical asteroid photometry, Gaia will not obtain full lightcurves, but sparse, single photometric data spread over five years. The number of detections depends on the orbits of the objects, being the average around 60--70 snapshots for main-belt asteroids. These data will cover a wide range of observational circumstances, and in particular wide ranges of ecliptic longitudes, resulting in a good coverage of aspect angle variation. In terms of observational cadence, these measurements will be similar to the data stored in the Minor Planet Lightcurve Database. Nonetheless, Gaia snapshots will be photometrically 10 times more accurate, and what is more important, homogeneous, in the sense that they will be measured by a single detector, and not by different telescopes. To put it in other words, these sparse data can be considered as the single points of a time-extended lightcurve, describing the photometric variation of the asteroids not over a single rotation period, but over five years, characterized by a continuous change of the observing circumstances. Actually, the inversion problem related with deriving physical parameters of asteroids from such measurements has become a topical issue, since not only Gaia, but also new ground-based survey telescopes such as the Large Synoptic Survey Telescope (LSST) will produce this kind of data.

The inversion technique specifically developed to invert the Gaia sparse data for asteroids \citep{cell06}, is based on a genetic algorithm, where the solution of the inversion problem is characterized by the best fit of a set of parameters that have been obtained by means of several random variations during a genetic mutation process. This solution should mitigate the risk of falling in secondary minima of the system and its capability to derive the \textit{correct} inversion solution has been shown in some experiments with Gaia simulated observations and also with real data collected during the ESA Hipparcos mission (see for instance \citealp{cell09} or \citealp{carbo12}). On the other hand, adding existing ground-based observations for a given asteroid is not speeding up the performance of this method (in fact the inversion becomes slower with greater number of measurements) and whether such observations can improve the method performance or not is a topic that needs to be studied. 

In this paper, we make a more general and detailed reassessment of the expected performances of the Gaia inversion algorithm. To do that, we fed the algorithm with simulations for tens of thousands of asteroids with different spin axis orientations, different rotational periods and random shapes. Such work is necessary to correctly analyse the results generated with the Gaia inversion algorithm at the end of the mission, when asteroids' photometric observations will be released.

Now that all the parameters of the Gaia scanning law are fixed, we are able to predict exactly the observation sequence for Solar system objects. It means that we can plan to observe from the ground at the same time as Gaia does. For example, we can very easily add a full rotational (dense) lightcurve around (or very close to) an isolated observation by Gaia. The link between the two data sets would then be very strong, as a single Gaia measurement provides a very precise absolute magnitude that can be used to calibrate the ground-based lightcurve. The question is: How many such lightcurves per object we need to obtain a substantial improvement of the inversion? Maybe a single one? Or more? Therefore, this work is thought to address such questions and lay the foundations for a collaboration involving coordinated observations from the ground.

\section{Control test with triaxial ellipsoids and "geometric" scattering law}

\begin{figure}
\centering
\includegraphics[width=80mm]{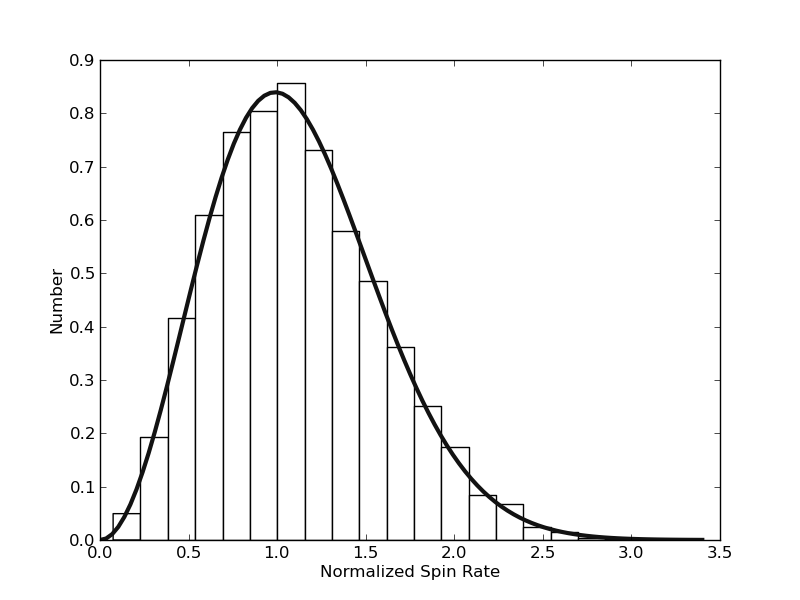}
\caption{Histogram of f/$\langle$f$\rangle$ for the asteroid population generated in our simulations.}
\label{period}
\end{figure}

The first test was performed to detect any systematic divergence between our simulated asteroids' magnitude and the magnitude generated by the Gaia inversion algorithm. Since the magnitude--phase relationship is essentially linear for the typical range of phase angles covered by Gaia \citep{zappala}, the inversion algorithm includes a linear parameter to describe this effect. Thus we have not implemented any light-scattering model in our simulations, but we have considered the geometrical phases \citep{lindegren}. For this test, we simulated Gaia-like observations for 10 359 triaxial ellipsoid shapes. This amount of objects is not a random choice, but the result of generating a set of asteroids having their spin axis directions uniformly distributed. The procedure followed to generate such uniform distribution start defining an initial mesh, consisting of eight unit vectors with respect to a common origin, each of those being the vertex of a cube. Then we recursively subdivide the surface with the Catmull--Clark subdivision method \citep{cat78}, which smooths the initial mesh surface by dividing the surface's polygons into smaller ones. After seven iterations we obtain a mesh with 10 359 vertices, each of those being the spin axis orientation of a given simulated asteroid. 

\begin{figure}
\centering
\includegraphics[width=80mm]{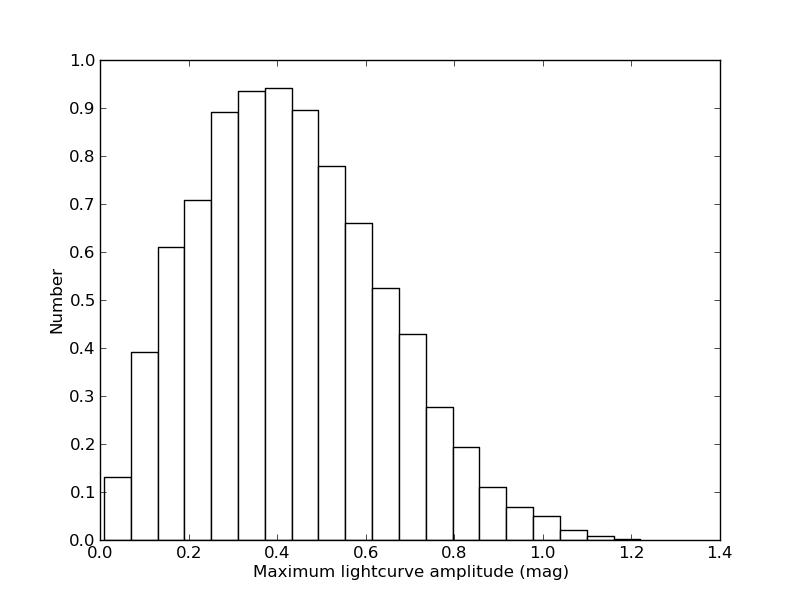}
\caption{Histogram of maximum lightcurve amplitude distribution for the asteroid population generated in our simulations.}
\label{amplitude}
\end{figure}

In order to generate the observational epochs for each object, we have used the Gaia mission simulator developed by F. Mignard and P. Tanga at the Observatoir de la C\^{o}te d'Azur (OCA), with a sample of asteroids having typical main-belt orbits. The population period distribution is shown in Fig.~\ref{period} and was generated following a Maxwellian distribution like the one described in \citet{pravec}. The disc-integrated photometry simulations have been generated using a Z-buffer standard graphic method described by \citet{catmull}. Z-buffering works by testing pixel depth. The z-value of any new point to be written into the buffer is compared with the z-value of the point already there. If the new point is behind, it is discarded, whereas if it is in front, it replaces the old value. We note that simpler and more efficient methods exist to generate the brightness of a triaxial ellipsoid, but the main objective of this test was to ensure that the algorithm used to generate the photometric simulations for further tests with more elaborate shape representations was well performing and we were not adding any bias in our analysis. The resulting distribution of magnitudes (i.e. the maximum lightcurve amplitude for each asteroid) is shown in Fig.~\ref{amplitude}.

\subsection{Test results overview: rotational period, spin axis orientation and overall shape}

The inversion run was executed using the Pozna{\'n}'s observatory cluster which consist of 27 workstations equipped with a six-core AMD processors (3 GHz), and the outcome was obtained after one full day of computations. In terms of pole determination, the results were positive, as the inversion algorithm found the correct pole (within 5 degrees of the true value) and shape (within 5 per cent of the true axis ratio) for more than 99 per cent of inversion runs. A few results presented an error in the pole determination, that increased as a function of the pole latitude. This situation can be interpreted as being caused by the double-pole ambiguity of derived spin states of asteroids orbiting close to the plane of the ecliptic. Thus this result is not an intrinsic problem of the method used, but a well-known limitation of the inversion techniques \citep{michalowski}.

\begin{figure}
\centering
\includegraphics[width=80mm]{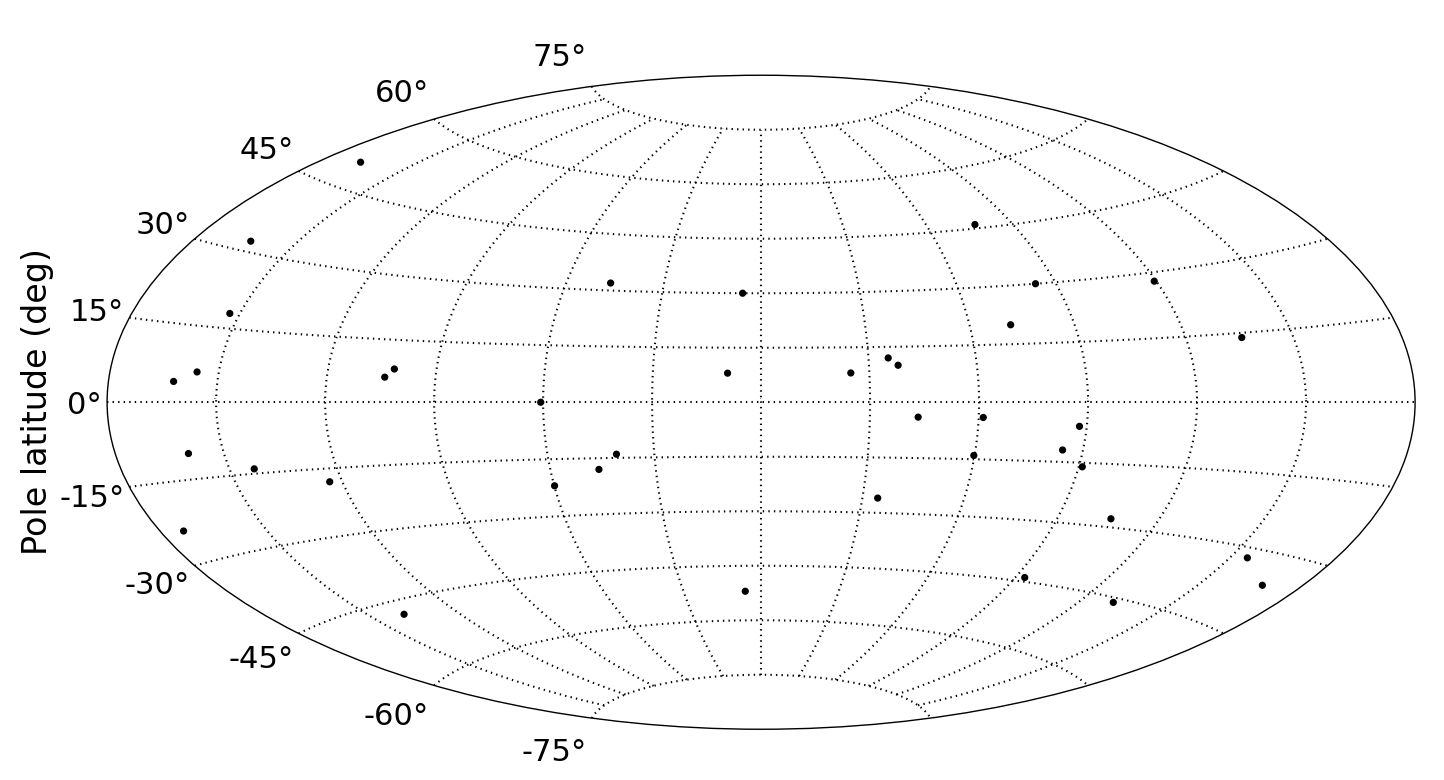}
\caption{Result of the period determination. The black points represent the original pole orientation for the runs which period determination had an error greater than 0.01 per cent. The number of wrong period determination was less than 0.4 per cent of the 10,359 asteroids used in this test.}
\label{okres}
\end{figure}

The correct rotational period was also found for the majority of inversion attempts. In particular, for 10 319 of the 10 359 runs the correct value was found with an accuracy better than 0.01 per cent. The original orientation of the spin axis (ecliptic longitude $\lambda$ and ecliptic latitude $\beta$) for the few attempts with a wrong period determination is shown in Fig.~\ref{okres}. The majority of the wrong solutions are confined to pole latitudes with $|\beta|<30$ degrees, and there was no wrong solution for pole latitudes higher than 60 degrees. As will be shown later, it is a standard behaviour of the method to better perform for asteroids having high pole latitudes. A similar behaviour has been found in the past for other inversion techniques (e.g. in \citealp{hanus11}). As for our case, this effect can be explained by understanding one of the major advantages (paradoxically) of Gaia's observations: its capability of seeing asteroids in a wide range of ecliptic longitudes in such a \textit{short} period of time (the operational mission phase is planned to last for five years). For instance, if we consider a main-belt asteroid having a pole with $\beta\sim0$ degrees, the aspect angle (i.e. orientation of the object's spin axis with respect to the direction of sight of the observer) will be very low for two out of five apparitions observed by Gaia (see Fig.2 in \citealp{cell06}). For such apparitions the asteroid's lightcurve is presenting almost no amplitude, resulting in the loss of information about the spin period signal in the sparse-in-time measurements obtained under such circumstances. If the observational sequence for such kind of objects is unluckily distributed, and the majority of measurements are collected under such geometries, the period search would become very sensitive to any asymmetries in the lightcurve, arising for example from an irregular shape. This could cause the genetic algorithm to find alternative solutions, resulting in a warning flag (refusing to generate a solution) or, in the worst case, it could lead to a wrong inversion solution. On the other hand, asteroids with high pole latitudes are always going to be observed under aspect angles far from zero (see fig.1 in \citealp{cell06}), thus in such cases, each measurement is bearing valuable information about the spin period. This is because, in such geometries, the instigator of the main periodical signal will be the asteroid axis with the greatest angular momentum (thus the longest). Consequently, the lightcurve amplitude will be near its maximum value and the signals due to any shape irregularity would play a secondary role. Nevertheless, it should be highlighted that the effect of low pole latitude plays in opposite ways for pole determination and for spin period determination. In particular, asteroids having low pole latitudes are ideal for any inversion technique for deriving the pole, due to the high variation in lightcurve amplitude for different ecliptic longitudes.

Concerning other results, we also studied the ellipsoid axis ratio b/a which describes the elongation of the body. This parameter was found with an accuracy better than 5 per cent for more than the 98 per cent of inversion attempts. In this case, no correlation with the pole latitude can be observed. On the other hand, in the case of the c/a axis ratio determination, there is a clear relation between its error and the pole latitude. In particular, the c/a axis ratio was found with an accuracy better than 5 per cent for 95 per cent of the attempts, and almost the totality of the problematic cases -- meaning solutions with an accuracy worse than 5 per cent -- were found for asteroids with extreme values of sin$\beta$. This result is not surprising and can be easily explained in terms of observational geometry, as for objects having high pole latitudes and orbits close to the ecliptic, the observations bring no or little information on the c axis.

Finally, it is worth mentioning that the presence of a small number of wrong inversion solutions is, in any case, unavoidable, due to the way the genetic approach works. There is always the possibility that several genetic inversion attempts for the same object will not be sufficient to catch the right solution. Increasing the number of attempts per object would certainly improve further the performances, but this is not feasible in the Gaia scenario due to CPU time constraints, as the inversion algorithm will have to process a number of the order of half a million of asteroids (in other words, two months of computations in our observatory cluster).

\begin{figure}
\centering
\includegraphics[width=80mm]{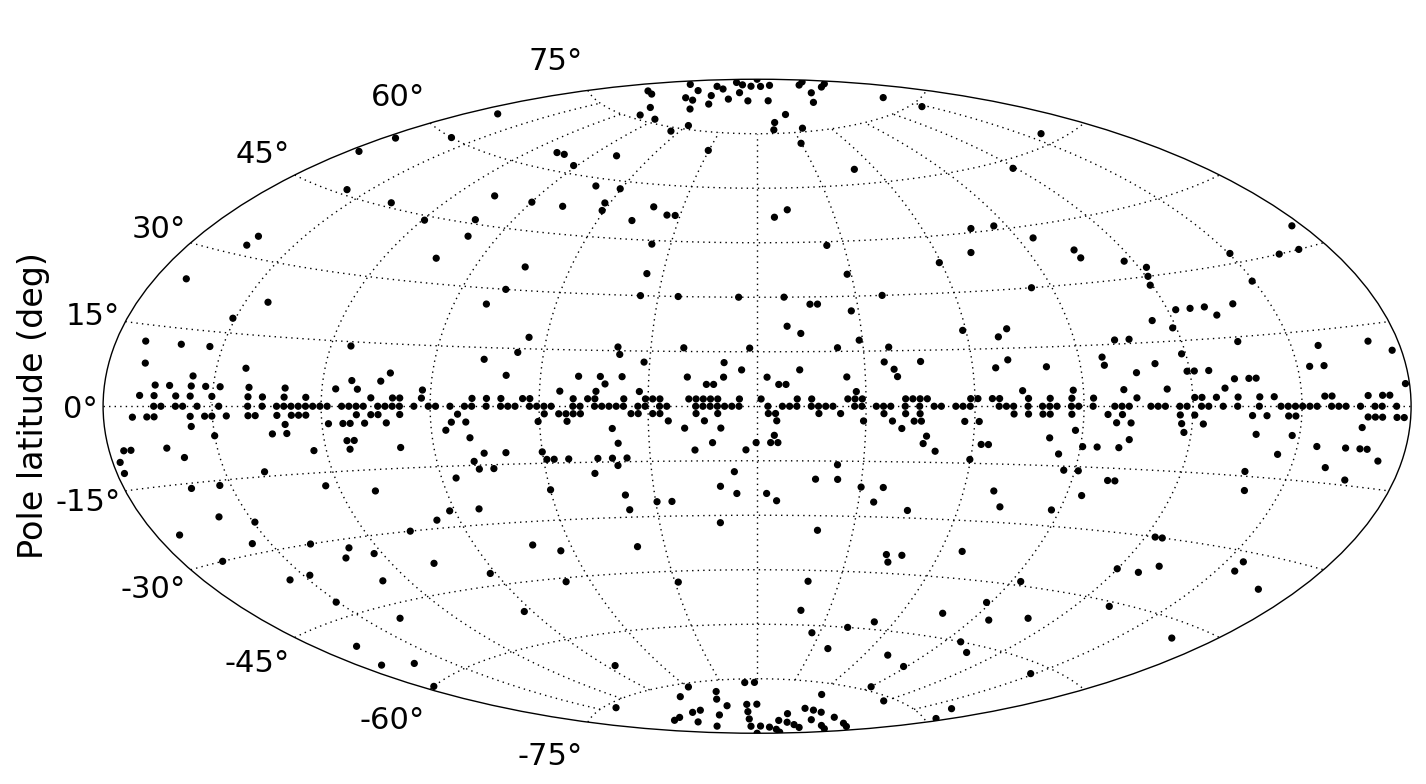}
\caption{Distribution of the warnings received from the inversion algorithm.}
\label{war}
\end{figure}

\subsection{Results control system}

Since \textit{dense} lightcurves are only available for a minority of the asteroids observed by Gaia, we would not be able to infer if the inversion results are providing real information of the asteroid's physical parameters or, in contrast, they are the result of an inversion artefact. In order to tackle this problem, the inversion algorithm developed for the Gaia data analysis pipeline, is including a \textit{warning} criterion to select the acceptable results. In particular, a \textit{warning} flag will be generated for those cases where the best fit is close to the second-best one, but their inversion solution is substantially different. It is still under discussion if these solutions shall be included in the final catalogue marked with a flag or, instead, they shall remain unpublished. The distribution of such cases can be seen in Fig.~\ref{war}. We received a total of 660 warnings, being 59 per cent of them from asteroids having a pole latitude between $-15<\beta<15$ degrees. This shall be taken into account when analysing the Gaia inversion statistics, as we expect them to show a lack of asteroids with low pole latitudes. Otherwise, one could tend to mislead this effect with some physical effects, such as non-gravitational forces.

\subsection{Simulations contaminated with Gaussian noise}

\begin{figure*}
\centering
\setlength\fboxsep{0pt}
\setlength\fboxrule{0pt}
\fbox{\includegraphics[width=150mm]{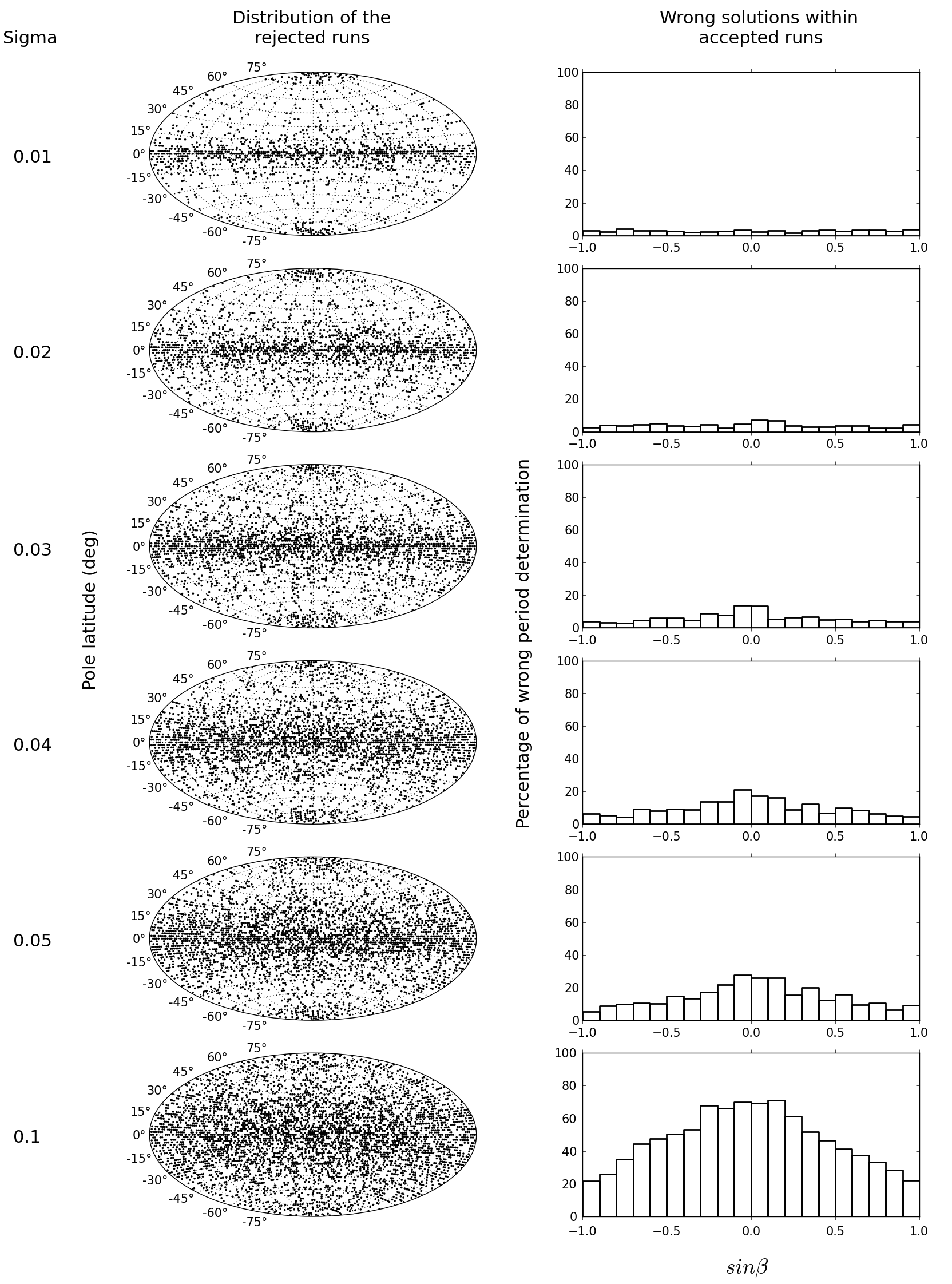}}
\caption{Distribution of the inversion results for each $\sigma$ value for the noise. The projections on the left show the inversion runs for which the solution was not accepted by the algorithm. The histograms on the right show (in per cent) the distribution of the wrong solutions within the accepted runs as a function of the asteroids' pole latitude.}
\label{gauss}
\end{figure*}

The good results obtained in the control test described above allow us to be confident with the methodology used, i.e. we are able to generate photometric simulations which are correctly inverted by the software algorithm that will be used for the analysis of Gaia asteroid photometry. However, we cannot expect the inversion of the real observations obtained by Gaia to have such a high reliability, as our simulations were generated under ideal assumptions (triaxial ellipsoid, geometric scattering law, no tumbling or binary asteroids, etc). Obviously this is not the situation we are going to face when analysing the Gaia photometry. Gaia photometric accuracy for each single transit will be of the order of 0.01 mag for objects as faint as $V=18.5$. Thus in the majority of cases the method systematic errors (coming, for instance, from the ellipsoid shape approximation or the scattering law used) will be of greater concern than the errors arising from the photometric accuracy. Thus, we contaminated our photometric simulations with Gaussian noise with different values of $\sigma$, and we repeated the inversion process for each case. The results distribution is shown in Fig.~\ref{gauss}, and two different biases can be observed: 1) population bias, 2) inversion reliability bias. The first one is connected with the warnings obtained from the results control system described above. The number of rejected solutions is not homogeneously distributed, as the majority of them are concentrated around asteroids having low latitude poles. The second bias is affecting the reliability of the obtained results. For $\sigma\geq0.03$ the results reliability is becoming proportional to the asteroid's pole latitude, being worse for the low latitude poles. Worth noting that the inversion solutions studied are the ones accepted by the algorithm's warning system, thus the first and the second bias are superimposed.

\section{Realistic test using random non-convex shapes}

\begin{figure}
\centering
\includegraphics[width=40mm]{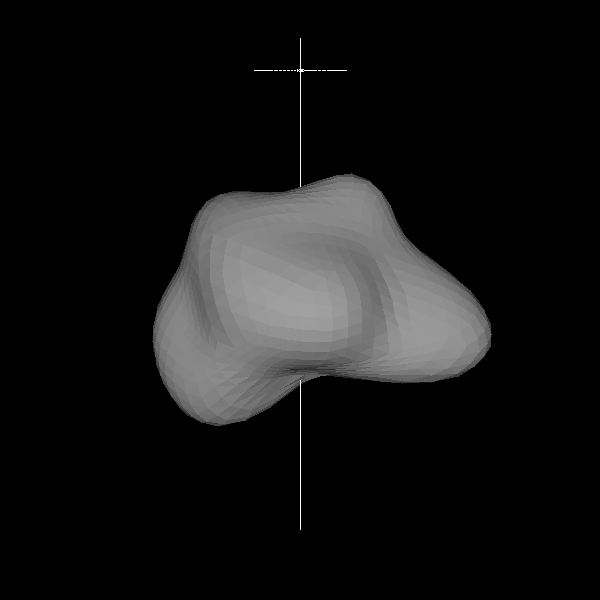}
\caption{Example of a random non-convex shape used for generating the photometric simulations.}
\label{ncshape}
\end{figure}

Once we felt fully confident with our simulation-inversion procedure and after studying the methodological bias, we proceeded with a more demanding test. In order to recreate as close as possible the kind of data which will feed the Gaia inversion algorithm we generated a set of 10 359 random non-convex shapes using Gaussian spheres \citep{muinonen}. An example of a random shape generated with our procedure can be seen in Fig.~\ref{ncshape}. The spin axis and the rotational period for each object were chosen following the same manner as in the previous tests described above. Next we generated the brightness using the Z-buffer standard graphic method. This method is including the phase angle effects, but unlike the first tests with ellipsoids, this is also including the shadowing effects produced by the irregular surface structures. Consequently, the lightcurves generated by these non-convex shapes presented complex features such as multiple minima and maxima, which cannot be recreated with a simple ellipsoid model. Moreover, once the brightness was simulated, we contaminated the set with a Gaussian noise with $\sigma = 0.03$. As stated before, this is three times the photometric accuracy expected for each single detection made by Gaia. But this might allow us to be on the safe side and would cover unexpected methodological errors, for instance, those connected with the scattering properties. Thus, the main question to be answered is whether the Gaia inversion algorithm will be able to deal with such complex scenario.  

\subsection{Results overview}

The answer to the previous question happened to be very optimistic: 65 per cent of the 10 359 asteroids were accepted by the inversion software and the correct solution was found for 83 per cent of them, or to be more precise, 6,754 inversion results were generated, from which 5 632 were correct and 1 122 incorrect. If we interpret this results in terms of the expected performance of the Gaia mission for Solar system objects, this would allow us to derive the poles, sidereal periods and axial ratios for several thousand asteroids.  Nonetheless, we could find some correlations between the majority of wrong inversions and certain parameters: the number of observations, the asteroid shape and the object's spin axis. Understanding such dependences would allow us to refuse some of the wrong results or even correct them.

\subsection{Influence of the number of measurements}

\begin{figure}
\centering
\includegraphics[width=85mm]{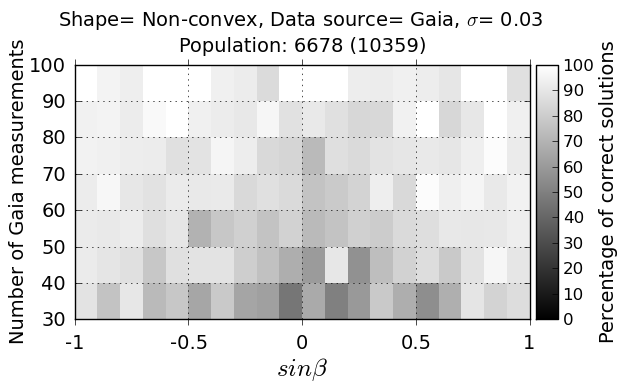}
\caption{Histogram showing the results obtained for the inversion of the simulated set of irregular body shapes. The percentage of correct solutions is plotted as a function of the number of Gaia detections for each bin of asteroid's pole latitude. The population number is indicating the amount of generated solutions and the total of inversion runs executed (in brackets).}
\label{nobs_nc}
\end{figure}

It is of common sense to consider that the more data points the inversion attempt has the better the result will be. And this is, in fact, what our results are confirming. The histogram in Fig.~\ref{nobs_nc} is showing the percentage of correct solutions found as a function of number of Gaia measurements and the asteroid pole latitude. The majority of regions having 70 measurements or more were found to be above the average of correct solutions. However, for lower regions, we found that the results are acceptable for asteroids having high pole latitudes, but worse than the average for asteroids having low pole latitudes. The range of latitudes around zero for which the inversion is presenting a lower reliability is getting wider the smaller the number of measurements is. Still, the good news is that, on the average, main-belt asteroids will be detected on the Gaia focal plane for a number of times between 60 and 70 during the five-year operational lifetime of the mission \citep{mignard}. Therefore the number of \textit{problematic} asteroids for which Gaia photometric inversion might produce wrong solutions will represent a small part (although still several hundreds) of the hundreds of thousands of asteroids observed.

\subsection{Influence of the asteroid shape}

The Gaia inversion algorithm is assuming the asteroids to have the shapes of triaxial ellipsoids. This approach was chosen mainly due to two reasons: 1) it seeks to minimize the CPU time required, 2) there was a need to produce an automated, standard procedure for working on such large amount of data in unattended runs. Although this approximation would seem inaccurate at a first glance, the results are showing that, despite its simplicity, this approach is sufficient to fit the data in the great majority of cases. Certainly, the shape solution provided by the algorithm is only a first-order approximation of the asteroid's shape and might provide only a general idea of the body elongation.

\begin{figure}
\centering
\includegraphics[width=70mm]{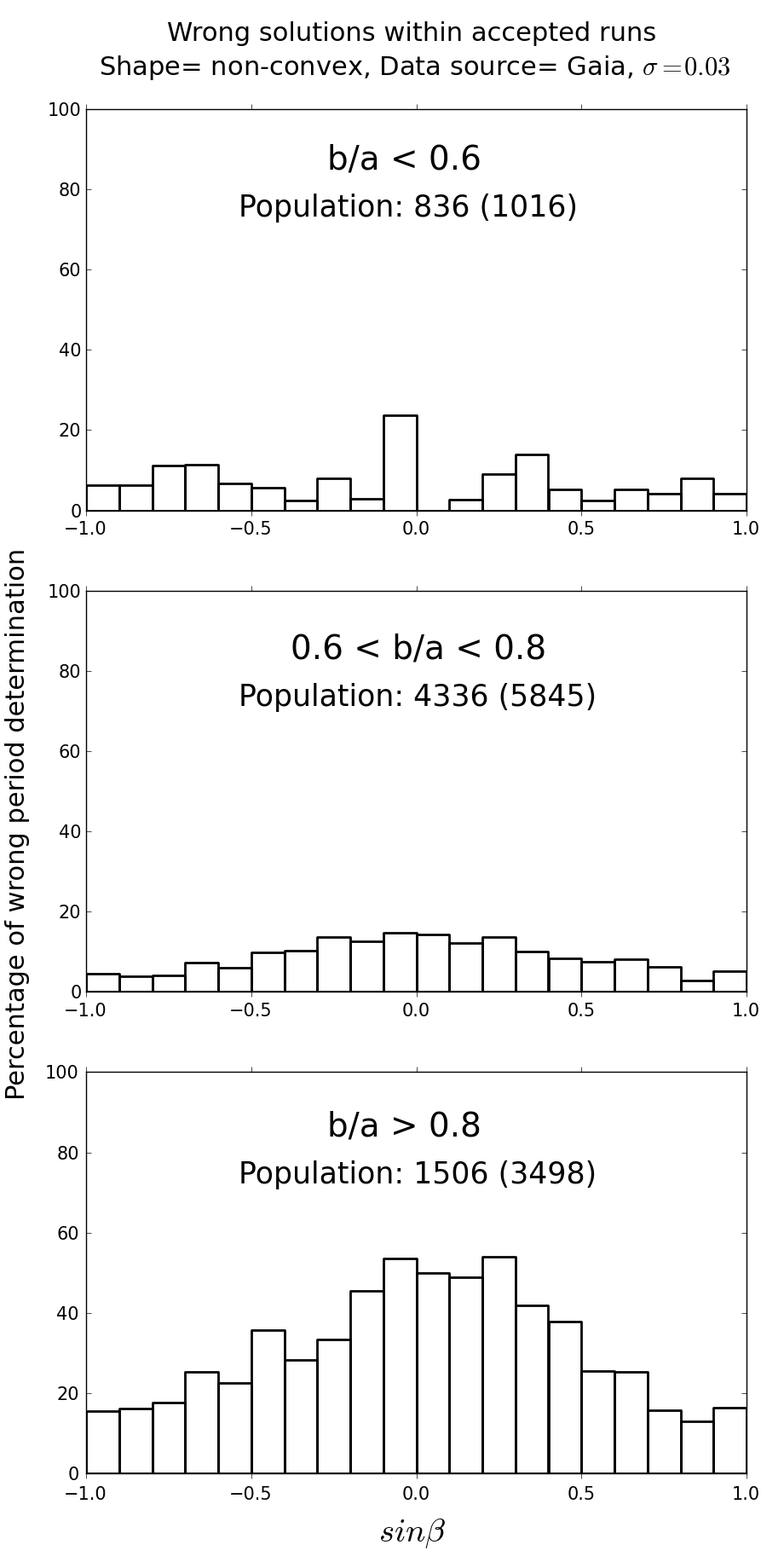}
\caption{Histograms showing the inversion results obtained for three different groups of asteroids as a function of their equivalent b/a axis ratio (see text). The population numbers are indicating the amount of generated solutions and the total of inversion runs executed (in brackets).}
\label{plot_ba}
\end{figure}

In order to assess the goodness of the inversion solution we calculated the principal moment of inertia for each random shape, and we determined the triaxial ellipsoid with an equivalent moment of inertia. This operation enables us to obtain an indicator of the elongation of any irregular shape, as the asteroids have been simulated in a relaxed state (i.e. with the rotation axis coincident with its principal axis of rotation). Finally, we classified our set of random shapes into three groups, according to the value of the equivalent b/a axis ratio calculated. The results are presented in Fig.~\ref{plot_ba}. As we could expect, the worst results are obtained for the quasi-spherical bodies and, in particular, for those having a low pole latitude. Such population is presenting the highest ratio of wrong solutions (around 30 per cent on the average) but also one out of two solutions is refused with a warning flag. These results are in agreement with the first tests presented above and can be explained using the same scheme (see Section 2.1 for more details). It is worth pointing out that, by definition, an ideally spherical object cannot be inverted, since the magnitude becomes dependent on the phase angle only. When b/a ratio approaches 1, the lightcurve amplitude remains always very small, and the inversion algorithm will find a large number of equivalent solutions (in terms of residuals) characterized by a large variety of possible poles.

\section{Gaia photometry combined with ground-based observations}

Combining Gaia observations of asteroids with ground-based lightcurves becomes straightforward when both observations are taken simultaneously. In contrast, if the lightcurve obtained from the ground does not include the epoch of observation by Gaia, there may be problems to link the Gaia observation to the rotational phase, and to calibrate the magnitudes of the ground-based data, especially in cases when the lightcurve is complex and the period resulting from the lightcurve is uncertain.

With the aim of supporting an observational campaign, it would be a good idea to publish the Gaia observation sequence for Solar system objects, allowing the observers to obtain a lightcurve of a certain asteroid at the same time as Gaia is collecting a very precise photometric measurement. Later on, it will be possible to calibrate the ground-based observation (even if it is relative photometry) with the Gaia absolute magnitude, and proceed with the inversion process normally. Formally, the only difference between data sources will appear during the preparation of the input file containing the photometric error associated with each observational instrument and the position vectors of the observer. 

In order to study the impact of adding ground-based observations to Gaia data, we have simulated a full lightcurve with 60 point measurements for the asteroids non-convex shapes described in Section 3. The particular geometry of the scan movement of Gaia telescopes, which never point on the Sun nor its opposition, results in observations taken at relatively high phase angles. For instance, considering the Gaia observations of a main-belt asteroid, the measurement with the lowest phase angle will be usually above 10 degrees. As the asteroid's magnitude becomes fainter for increasing phase angles, we selected the date of the Gaia's measurement with the lowest phase angle as an epoch to generate the lightcurve. This choice was taking into account that ground-based supporting observations of asteroids will be probably done by small or mid-sized telescopes, and moreover, we are not interested in the projecting shadows appearing at higher phase angles. Finally, we contaminated all the simulated lightcurves with Gaussian noise ($\sigma=0.03$). It should be pointed out that the results presented here are limited to the specific choice made when adding dense lightcurves and they could be different for observations obtained around opposition, more than one curve, etc.

\subsection{Impact on the results as a function of the asteroid shape}

\begin{figure*}
\centering
\setlength\fboxsep{0pt}
\setlength\fboxrule{0pt}
\fbox{\includegraphics[width=150mm]{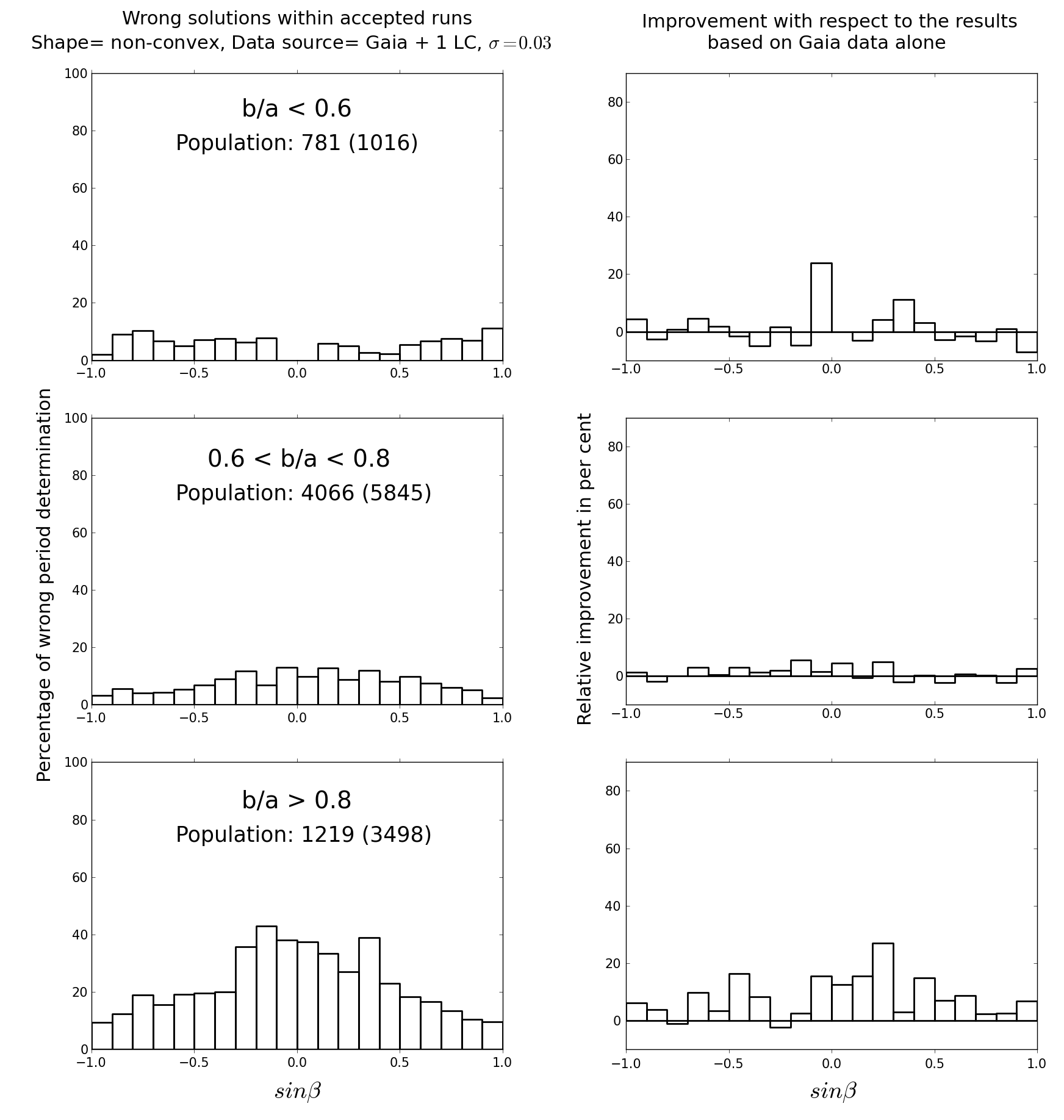}}
\caption{Histograms on the left show the inversion results obtained for a data set combining Gaia photometry and one full lightcurve. The results are divided into three groups as a function of the asteroids' equivalent b/a axis ratio and are plotted as a function of the asteroids' pole latitude. Histograms on the right show the relative improvement comparing with the inversion results obtained for Gaia data alone.}
\label{impro_ba}
\end{figure*}

After combining the simulated Gaia photometry and the full lightcurve, we use the resulting data set to feed the inversion algorithm. The results resemble the ones obtained for the Gaia data alone, i.e. a good overall result, albeit worse reliability for nearly spherical bodies. In Fig.~\ref{impro_ba} we show the percentage of wrong solutions as a function of the asteroid's pole latitude and the equivalent b/a ratio, as well as the improvement of combining both data sets. It resulted that, with the exception of nearly spherical objects, the improvement was almost negligible. It is worth noting that the relative improvement on the results is not caused by an increment of the correct inversions, but the reduction of the accepted wrong solutions. 

\subsection{Impact on the results as a function of the number of measurements}

\begin{figure}
\centering
\includegraphics[width=80mm]{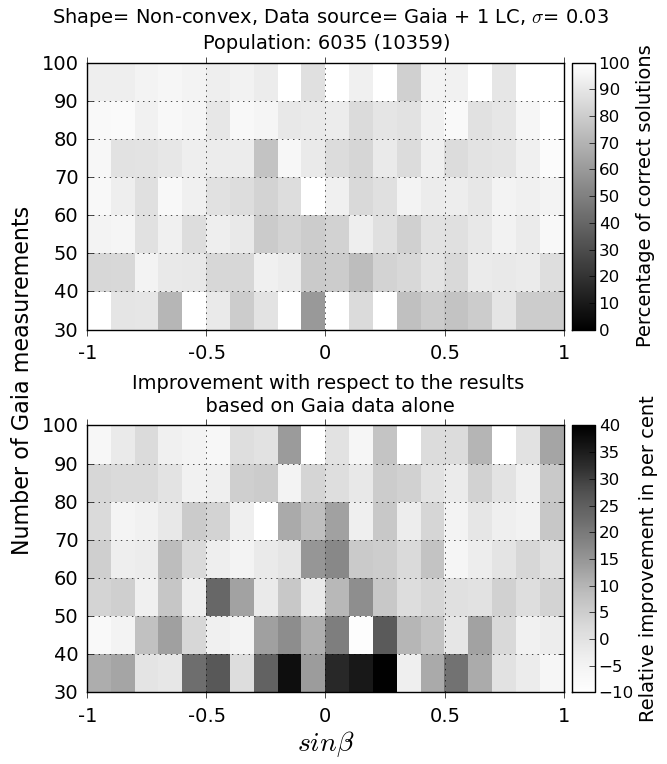}
\caption{The histogram on the top shows the results obtained for the combined data set. The percentage of correct solutions is plotted as a function of the number of Gaia detections for each bin of asteroids' pole latitude. The population number is indicating the amount of generated solutions and the total of inversion runs executed (in brackets). The histogram on the bottom shows the relative improvement compared to the inversion results obtained for Gaia data alone.}
\label{histo_nc_1lc}
\end{figure}

If we analyse the results as a function of the amount of points collected by Gaia, the improvement can be clearly appreciated for asteroids having less than 50 detections. The results of the inversion run and the corresponding improvement are shown in Fig.~\ref{histo_nc_1lc}. For the majority of asteroids for which supporting ground-based observations will be planned, we would not have any a priori information of the physical parameters. In contrast, the total number of observations for each asteroid can be already calculated as all the parameters of the Gaia scanning law are already fixed and the mission has started its science phase. Considering that it is not feasible to obtain one lightcurve for each of the $\sim$300 000 asteroids observed by Gaia, it would be necessary to draw up an observational plan, thus we conclude that, for the purposes, the number of expected Gaia measurements can be a good selection criterion.

\subsection{Discussion of the results}

The results presented above could seem counter-intuitive. In particular, one could argue how it is possible that the addition of a full lightcurve does not generally improve the reliability of the method. In order to understand the situation, we should clarify some points:

\begin{enumerate}
\item \textit{The actual version of the inversion algorithm treats equally each single measurement.} 

The goodness of the inversion solution is estimated on the basis of the fit between computed and observed single measurements. In the case of asteroids with abundant Gaia observations (for instance, more than 80 points), a single lightcurve will have a discreet influence on the inversion result, especially for lightcurves with low amplitude \citep{ania}. This situation can be faced by increasing the weight of the ground-based lightcurve to the detriment of the Gaia data. 

\item \textit{The additional lightcurves were blindly simulated in terms of asteroid's aspect angle.} 

Observations obtained at high aspect angles represents the best-case scenario when deriving the rotation period, as asteroid lightcurve is then at its maximum amplitude. However, it would require to know in advance the spin axis orientation of the given asteroid so to predict the appropriate observational epoch. For the great majority of asteroids observed by Gaia this is not feasible, as we do not know their rotational states. For this reason, the only selection criterion used when generating the additional lightcurves was the asteroid's phase angle (see the beginning of the section), thus some of the lightcurves present low amplitudes. Under this particular situation, the inversion fit's residual is very low no matter the rotation period, and so the impact on the inversion results is negligible. 

\item \textit{The triaxial ellipsoid assumption of the inversion algorithm might have not well-behaved cases.} 

While it has been proven to work well for the majority of cases, the triaxial ellipsoid assumption might cause the inversion to fail under tough cases. For instance, very irregular shapes can generate lightcurves with multiple extrema, which cannot be inverted using a triaxial ellipsoid assumption, no matter how many complementary data are used.   
\end{enumerate}
 
\section{Conclusions}

We have tested the Gaia inversion algorithm fed with realistic simulations of asteroids and the results have been encouraging. The number of correct inversions remains above 80 per cent even under severe scenarios with photometric errors of 0.03 mag. Moreover, we have detected the most problematic scenarios for the method: 1) asteroids having a quasi-spherical shape, 2) asteroids with low pole latitudes and 3) asteroids with less than 50 data points. These biases have to be considered before sketching physical interpretations of the future inversion results.
Otherwise, the detected bias could be erroneously mislead with physical effects, like an overinterpretation of the YORP effect resulting from the loss of many cases having poles far from perpendicular to the orbital plane. We have shown that it is possible to reduce the number of wrong solutions by adding a single lightcurve to Gaia's measurements. Thus this pre-selection method can be used to coordinate an observational campaign with the aim of enhancing the Gaia Solar system science output. It is also of utmost importance to develop strategies for collaboration with ground-based optical surveys that will produce in the near future sparse photometric measurements similar to the ones produced by Gaia in terms of quantity and quality. The most outstanding project for the next decade is LSST, which, up to some extent, could be understood as a ground extension of the Gaia mission. In this sense, the inversion algorithm used in this paper can be easily adapted to process and combine data from other surveys. Such collaboration shall even greatly boost our statistical picture of physical parameters from Solar system objects and would allow us to derive shapes and spin states of asteroids far beyond the main belt.

\section*{Acknowledgements}

We thank F. Mignard and Ch. Ordenovich (OCA, Nice) for putting at our disposal the use of the Gaia simulator of Solar system observations.

The work of TSR was carried out through the Gaia Research for European  Astronomy  Training  (GREAT-ITN)  network.  He  received  funding  from  the  European  Union  Seventh Framework Programme (FP7/2007-2013) under grant agreement no. 264895.

The work of AC was partly funded by ASI contract I/058/10/0.

\label{lastpage}

\end{document}